\documentstyle[11pt,paspconf,epsf]{article}

\begin{document}

\title{Microwave Emission from Galactic Dust Grains}
\author{B. T. Draine}
\affil{Princeton University Observatory, Princeton University,
	Princeton, NJ 08544-1001, USA; {\tt draine@astro.princeton.edu}}
\author{A. Lazarian}
\affil{Canadian Institute of Theoretical Astrophysics,
	60 St. George St., Toronto, ON M5S 3H8, Canada;
	{\tt lazarian@cita.utoronto.edu}}

\begin{abstract}
Observations of the cosmic microwave background have revealed a
component of 10--60~GHz
emission from the Galaxy which correlates with 100--140$\mu$m
emission from interstellar
dust but has an intensity much greater than expected for
the low-frequency tail of the ``electric dipole vibrational''
emission peaking at $\sim$130$\mu$m.
This ``anomalous emission'' is more than can be accounted for
by dust-correlated free-free emission.
The anomalous emission could be due in part to magnetic dipole emission from
thermal fluctuations of the magnetization within interstellar dust grains, but
only if a substantial fraction of the Fe in interstellar dust resides
in magnetic materials such as metallic iron or magnetite.
The observed anomalous emission is probably due primarily to
electric dipole radiation
from spinning ultrasmall interstellar dust grains.
This rotational emission is expected to be partially polarized.
\end{abstract}


\section{Introduction \label{sec:intro}}

Microwave emission from interstellar dust grains contributes a
``galactic foreground'' which must be recognized and subtracted from
observations of the sky in order
to carry out studies of angular structure in 
the cosmic microwave background radiation (CMBR).
This microwave emission also offers us new information on
the properties of interstellar dust grains.

In \S\ref{sec:obs} we review the observational evidence for 
infrared and microwave emission from interstellar dust.
There are several sources of ``galactic foreground'' emission
at microwave frequencies: relativistic electrons (synchrotron emission),
thermal electrons (free-free emission),
and dust grains. 
Microwave emission from dust grains was discovered by studies of the
CMBR, which revealed a surprisingly strong
component of the microwave sky brightness which was correlated with
interstellar matter, as traced by 100$\mu$m thermal dust emission.
Because this dust-correlated signal was much stronger than expected
from existing models of interstellar dust, it has been referred to
as ``anomalous'' emission (Leitch {\it et al.}\ 1997; Kogut 1999).

The spectrum of the ``anomalous'' microwave emission is not consistent with
synchrotron emission, and maps at 408 MHz (Haslam 1981) and 
1.42 GHz (Reich \& Reich 1988)
do not correlate with the observed 15-100 GHz intensity
(Kogut {\it et al.}\ 1996a,b;
Leitch {\it et al.}\ 1997;
de Oliveira-Costa {\it et al.}\ 1997, 1998),
so the anomalous emission is evidently not synchrotron radiation from
relativistic electrons.
In section \ref{sec:free-free} we conclude that
the observed emission significantly exceeds the free-free emission from
interstellar plasma.  The excess emission must be due to dust.

There are three quite distinct mechanisms whereby dust can 
radiate
at microwave frequencies; they can be classified as
(1) ``vibrational electric dipole'' emision (due to thermal fluctuations
in the charge distribution in the grain); (2) ``magnetic dipole'' emission
(due to thermal fluctuations in the magnetization of grain material);
and (3) ``rotational electric dipole'' emission (due to the rotating
electric dipole moment of a spinning grain).

Most of the power radiated by interstellar dust is due to
``vibrational electric dipole'' emission, peaking in the far-infrared
at $\sim$100$\mu$m.
The low-frequency ``tail'' of this emission
can be extrapolated to microwave frequencies, but
falls far below the observed 10--60 GHz emission.
However, as discussed in \S\ref{sec:magdipole},
if grains contain magnetic materials -- such as magnetite or
metallic iron -- the thermal fluctuations in the magnetization will
result in strong 
magnetic dipole emission.

There is strong independent evidence for a large population of ultrasmall
grains.
These grains should be spinning rapidly, should have electric
dipole moments, and therefore should radiate at microwave frequencies,
as discussed in \S\ref{sec:rotational}
The expected spatial variations of the microwave emission from dust
are discussed in \S\ref{sec:space}
The possibility that this radiation can be polarized is discussed in
\S\ref{sec:polarization}, and observational tests are considered in
\S\ref{sec:tests}
We summarize in \S\ref{sec:summary}

\section{Observations \label{sec:obs}}

Figure \ref{fig:obs} shows the observed emission spectrum
of diffuse interstellar dust, based on measurements by the
{\it InfraRed Astronomy Satellite (IRAS)},
the {\it FIRAS} spectrometer and the
{\it DIRBE} photometers on the {\it COsmic Background Explorer (COBE)},
and the {\it Mid-IR Spectrometer (MIRS)} and the
{\it Near-IR Spectrometer (NIRS)} on the
{\it InfraRed Telescope in Space (IRTS)}.

\subsection{Far-Infrared Emission \label{sec:fir}}
The emission from interstellar matter between
1~mm (300 GHz)
and $100\mu$m (3000 GHz) is due primarily to thermal emission from
dust particles heated by diffuse starlight to 
temperatures $T_{\rm d}\approx 15-25$K, as was
expected theoretically (see, e.g., Draine \& Lee 1984).

While the observed spectrum between 1~mm and 100$\mu$m 
can be accurately described by multicomponent fits 
(Wright {\it et al.}\ 1991;
Reach {\it et al.}\ 1995;
Finkbeiner \& Schlegel 1999;
Finkbeiner, Schlegel, \& Davis 1999)
the observed 1~mm -- 100$\mu$m emission,
with $\nu j_\nu$ peaking at $\lambda_p\approx130\mu{\rm m}$, can 
be approximated quite well by emission from dust 
with absorption cross section $\propto \nu^\beta$
and a single temperature $T_{\rm d}\approx hc/(4+\beta)\lambda_p k$.

\begin{figure}
\plotone{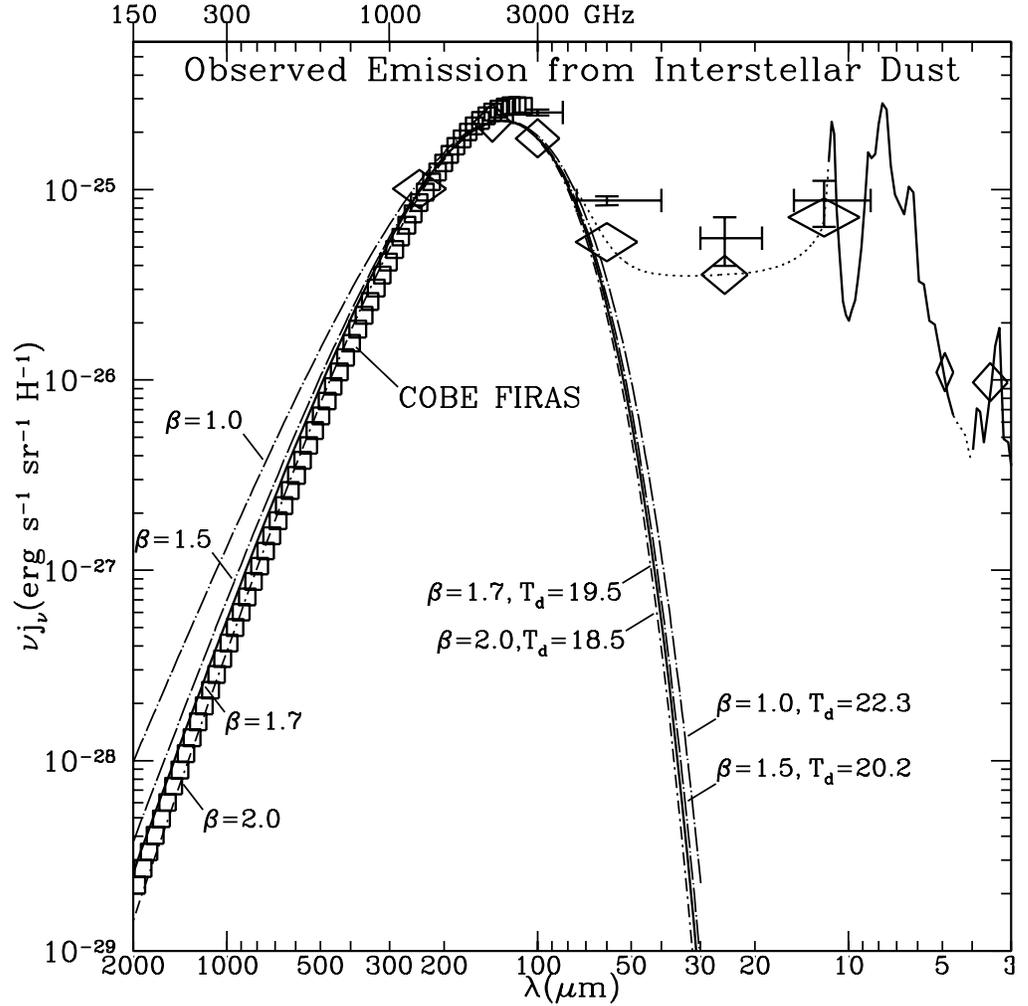}
\caption{\small%
	Infrared emissivity per H, based on data from
	{\it IRAS} (crosses; Boulanger \& Perault 1988),
	{\it COBE-FIRAS} (Wright {\it et al.}\ 1991);
	{\it COBE-DIRBE} (diamonds: Arendt {\it et al.}\ 1998);
	and
	{\it IRTS} (heavy curve: Onaka {\it et al.}\ 1996,
	Tanaka {\it et al.}\ 1996).
	An assumed stellar continuum has been subtracted from
	the {\it IRTS} spectra.
	The dotted line is to guide the eye.
	The {\it IRTS} spectra show emission features at 3.3,
	6.2, 7.7, 8.6, and 11.3$\mu$m.
	}
\label{fig:obs}
\end{figure}
Away from molecular regions, the 100$\mu$m emission correlates very
well with 21 cm emission (Boulanger \& P\'erault 1988), allowing us to
infer the average emissivity per H nucleon (Draine 1999):
\begin{equation}
\label{eq:betafit}
\nu j_\nu = \lambda j_\lambda \approx 2.3\times10^{-25}
\left(\frac{130\mu{\rm m}}{\lambda}\right)^{4+\beta}
\left[\frac{\exp(4+\beta)-1}{\exp(h\nu/kT_{\rm d})-1}\right]
{\rm erg~s^{-1} sr^{-1} H^{-1}}
\end{equation}
where a good fit is obtained for $T_{\rm d}=19.5$K and $\beta=1.7$.
This choice of $\beta$ and $T_{\rm d}$ 
corresponds to an absorption cross section per H nucleon
\begin{equation}
\frac{\tau_\nu}{N_{\rm H}} = 2.5\times 10^{-25}
\left(\frac{100\mu{\rm m}}{\lambda}\right)^{1.7} {\rm cm^2 ~H^{-1}} ~~~.
\end{equation}
Figure \ref{fig:obs} shows this and other single-temperature fits 
together with
observational determinations of the dust emission spectrum.
\subsection{Mid-IR Emission \label{subsec:midir}}
A striking feature of Figure \ref{fig:obs} 
is the strong emission at $\lambda < 50\mu$m observed by
{\it IRAS}, {\it COBE-DIRBE}, and {\it IRTS}.
This emission, accounting for $\sim$35\% of the total power radiated
by dust, is far in excess of what would be expected from
dust grains at $T\approx20$K.
The only natural explanation for this is to attribute it to
dust grains which are so small that absorption of a
single starlight photon heats them to temperatures high enough for
thermal emission to produce the observed radiation
(Draine \& Anderson 1985).
Grain temperatures as high as $\sim200$K are required to account for
the emission observed in the 12$\mu$m band, so the grains must
be {\it very} small: if the specific heat is characterized by a Debye
temperature $\Theta=420$K (the value for graphite; 
Furukawa {\it et al.}\ 1972),
then a 6 eV 
starlight photon could heat a grain with $N\approx 280$ atoms to
$T=200$K.
Even smaller particles are required by the strong emission observed
at shorter wavelengths (see Figure \ref{fig:obs}).
The population
of small particles must be {\it very} large, since they must account
for $\sim$35\% of the total absorption of energy from starlight!

\subsection{Extrapolation to Lower Frequencies\label{subsec:extrap}}
Since eq.\ (\ref{eq:betafit}) fits the 1~mm -- 100$\mu$m emission so
well, it is natural to extrapolate it to lower frequencies.
A good single-temperature fit is for $\beta\approx1.7$, close to the
value $\beta=2$ predicted by
simple models for the response of dielectrics at frequencies below
all the optically-active resonances as well as simple models of conducting
materials (Draine \& Lee 1984).
\begin{figure}
\plotone{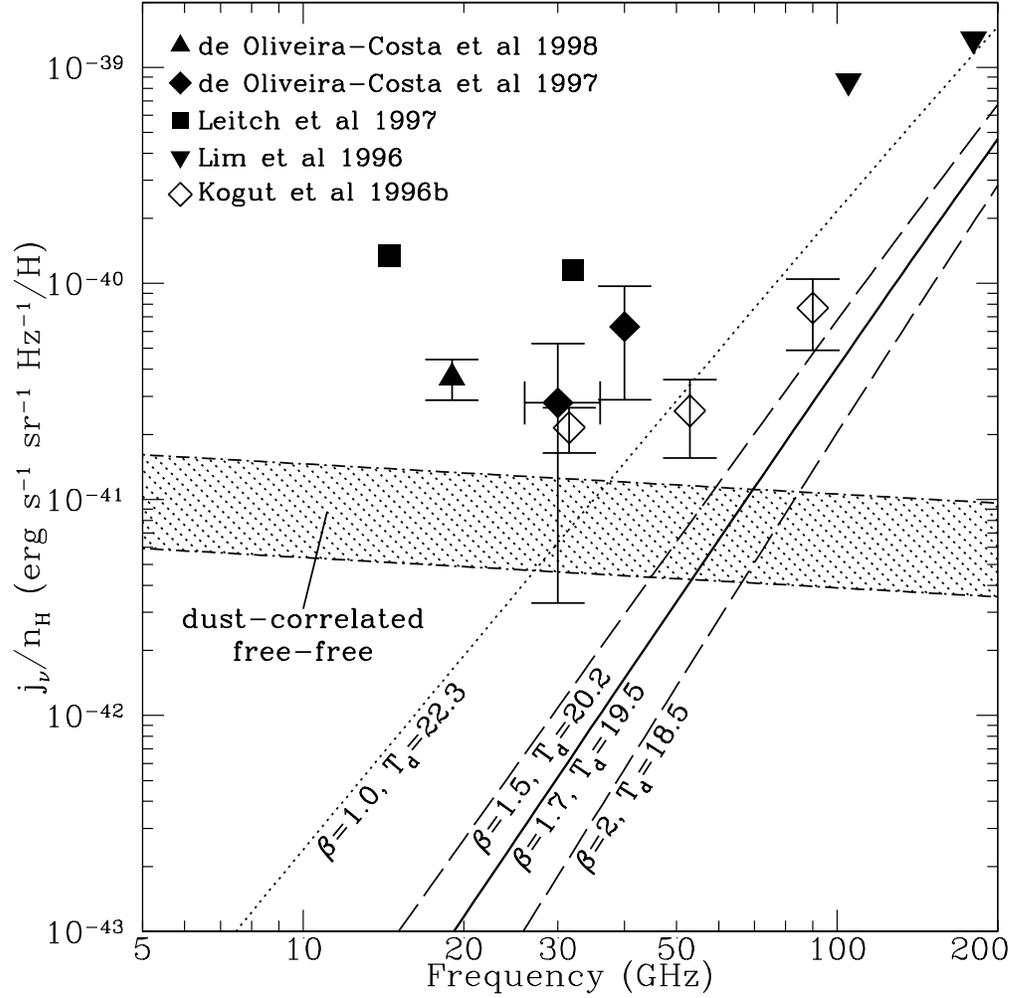}
\caption{\small%
	Emissivity per H at $\nu < 200$GHz,
	together with prediction of
	eq.\ ({\protect\ref{eq:betafit}}) for various choices of
	$\beta$.
	``Observed'' emissivities shown here and in following figures
	are based on observations of microwave emission correlated with
	100--140$\mu$m emission.
	In regions with cooler dust, the emissivity shown here will
	overestimate the actual emissivity (see text); this
	might explain the relatively high emissivities obtained from
	the observations of Leitch {\it et al.}\ and Lim {\it et al.}
	The dust-correlated H$\alpha$ emission (see
	{\protect \S \ref{sec:free-free}}) has been used to
	estimate the dust-correlated free-free emission; the shaded
	region corresponds to the uncertainties given in 
	eq.\ ({\protect\ref{eq:halpha}}).
	The observed emission at $\nu < 60$ GHz substantially exceeds
	both the dust-correlated free-free emission and
	the predicted thermal emission from dust with $\beta\approx1.7$.
	}
\label{fig:microwave_obs}
\end{figure}

In Figure \ref{fig:microwave_obs} 
we show the prediction for $\beta=1.7$ and three other values
of $\beta$. 
Note that the $\beta=1$ extrapolation
overestimates the emission for $2000 > \lambda > 300\mu$m
(see Fig.\ \ref{fig:obs}), and is
included in Figure 
\ref{fig:microwave_obs} only as an extreme example.

Figure \ref{fig:microwave_obs} also shows the
observed dust-correlated emissivity $j_\nu$ 
per H nucleon deduced for
diffuse interstellar matter
by cross-correlating the
measured microwave sky brightness $I_\nu$ with an FIR sky map 
(either {\it DIRBE} 140$\mu$m
or {\it IRAS} 100$\mu$m).
The best-fit slope $\Delta I_\nu/\Delta I_\nu({\rm FIR})$ 
reported by different experiments
is then converted to an emissivity
per H nucleon using the $140\mu$m or $100\mu$m emissivity from
eq.\ (\ref{eq:betafit}).

Equation \ (\ref{eq:betafit}) is appropriate only for dust in diffuse
regions.
The procedure used here 
will tend to {\it overestimate} the emissivity
$j_\nu = I_\nu/N_{\rm H}$ when
the observations include regions (e.g., non-star-forming molecular
clouds) 
where the dust is somewhat cooler than in diffuse HI clouds.
The cooler dust will have a lower FIR emissivity, and 
hence $N_{\rm H}$ will actually
be larger than deduced from $I_\nu({\rm FIR})$ using eq.\
(\ref{eq:betafit}).
This might explain the relatively high 
emissivities shown in Fig.\ \ref{fig:microwave_obs}
obtained from the observations
of Leitch {\it et al.}\ (1997),
as the fields observed by these experiments could contain significant
amounts of molecular gas 
(Finkbeiner, private communication).

The much larger fields observed by Kogut {\it et al.}\ (1996b) and
de Oliveira-Costa {\it et al.}\ (1997,1998) are expected to be
dominated by diffuse atomic gas, so the emissivities deduced from these
experiments are probably more accurate.
Better estimates of emissivities $j_\nu = I_\nu/N_{\rm H}$ 
could be obtained if the
H column density $N_{\rm H}$ is estimated using
the maps of dust optical depth produced by Schlegel {\it et al.}\
(1998), who used the {\it DIRBE} 100 and 240$\mu$m maps to allow for
dust temperature variations.

The observed dust-correlated 
microwave emissivities at $\nu < 60$ GHz tend to be well
above the extrapolation to low frequencies with the
``best fit'' $\beta\approx1.7$.
For this reason the dust-correlated microwave emission has been
referred to as ``anomalous'' (Leitch {\it et al.} 1997).

\section{Free-Free Emission? \label{sec:free-free}}

Since the observed dust-correlated 
emission at $\nu < 60$GHz is so much greater than
expected from simple extrapolation of the dust spectrum 
from far-infrared and sub-mm wavelengths,
we must consider what other mechanisms could be responsible.
Kogut {\it et al.}\ (1996a) suggested that the observed 30--50 GHz excess
was due to free-free emission from ionized gas which was correlated with dust.
However, Leitch {\it et al.}\ (1997) measured the dust-correlated 
14.5 and 32 GHz emission
from 36 fields around the North Celestial Pole, and showed that for these
fields the ``anomalous emission'' could not be free-free emission from gas at
$T_{\rm gas}\leq10^4$K, as the recombination radiation which must accompany
free-free emission is not present on H$\alpha$ maps
(Gaustad {\it et al.}\ 1996).

Leitch {\it et al.}\ 
noted that if the anomalous emission in this region were instead 
free-free radiation from 
plasma with $T_{\rm gas} > 10^6$K, the predicted
H$\alpha$ would be consistent with existing limits.
However, Finkbeiner \& Schlegel (1999) have recently reported
a {\it negative} correlation between the Leitch {\it et al.}\ 14.5 GHz
measurements and emission in the ROSAT X-ray C Band, so it is not
clear that hot gas can explain the anomalous emission in the
Leitch {\it et al.}\ fields.

McCullough {\it et al.}\ (1999) review H$\alpha$ observations and the 
contribution of free-free emission
to the Galactic microwave foreground.
Observing programs now underway will soon provide a much better knowledge
of the H$\alpha$ sky,
but existing studies of the correlation between H$\alpha$ emission and dust
100$\mu$m emission (McCullough 1997; Kogut 1997) over limited regions
find a dust-correlated H$\alpha$ component with
\begin{equation}
\frac{I({\rm H}\alpha)}{I_{100}}\approx (0.65\pm0.30) 
\frac{{\rm Rayleigh}}{{\rm MJy\,sr^{-1}}} ~~.
\label{eq:halpha}
\end{equation}
This correlation is found when averaging on angular scales of a few
degrees (McCullough {\it et al.}\ 1999).
The dust-correlated H$\alpha$ presumably originates mainly from photoionized
cloud rims, but approximately 25\% of the high latitude H$\alpha$ is
estimated to be scattered light (Jura 1979; McCullough 1997).
Thus we take the H$\alpha$ {\it emission} correlated with dust to be
75\% of eq.\ (\ref{eq:halpha}):
\begin{equation}
\frac{I({\rm H}\alpha\,{\rm em})}{I_{100}}\approx (0.49\pm0.23) 
\frac{{\rm Rayleigh}}{{\rm MJy\,sr^{-1}}} ~~.
\label{eq:halphaem}
\end{equation}
In Figure \ref{fig:microwave_obs} we show the dust-correlated 
free-free emissivity which
would accompany the dust-correlated H$\alpha$ emission from
eq.\ (\ref{eq:halphaem}), assuming
$T_{\rm gas}\approx 8000$K.
Figure \ref{fig:microwave_obs} shows
that the observed dust-correlated 10--60 GHz emission is systematically
well above the level expected for free-free emission 
from $\sim$10$^4$K gas.
An additional source of emission is required.\footnote{%
	McCullough {\it et al.}\ argue that the observational uncertainties 
	are still large enough that one cannot conclusively reject the 
	hypothesis that the 
	observed microwave excess is due to free-free emission 
	from $\sim$10$^4$K gas, but we feel that
	the evidence for an additional source of dust-correlated
	microwave emission -- as seen in 
	Fig.\ {\protect\ref{fig:microwave_obs}} -- is very strong.
	See also Kogut (1999).
	}

Hot gas cannot explain the emission observed over large fields by
Kogut {\it et al.}\ (1996) and de Oliveira-Costa {\it et al.}\ (1997, 1998),
since the resulting X-ray power would exceed the estimated 
rate of energy input from
supernovae by a factor $\sim$10$^2$ (Draine \& Lazarian 1998a,
hereafter DL98a).

Since synchrotron radiation has been ruled out, 
dust appears to be the only possible source of the excess emission.
Since the predicted ``vibrational electric dipole'' emission has been
seen above to be insufficient,
one or more additional dust emission mechanisms must be important at microwave
frequencies.

\section{Magnetic Dipole Emission? Possibly.\label{sec:magdipole}}

\begin{figure}[t]
\plotone{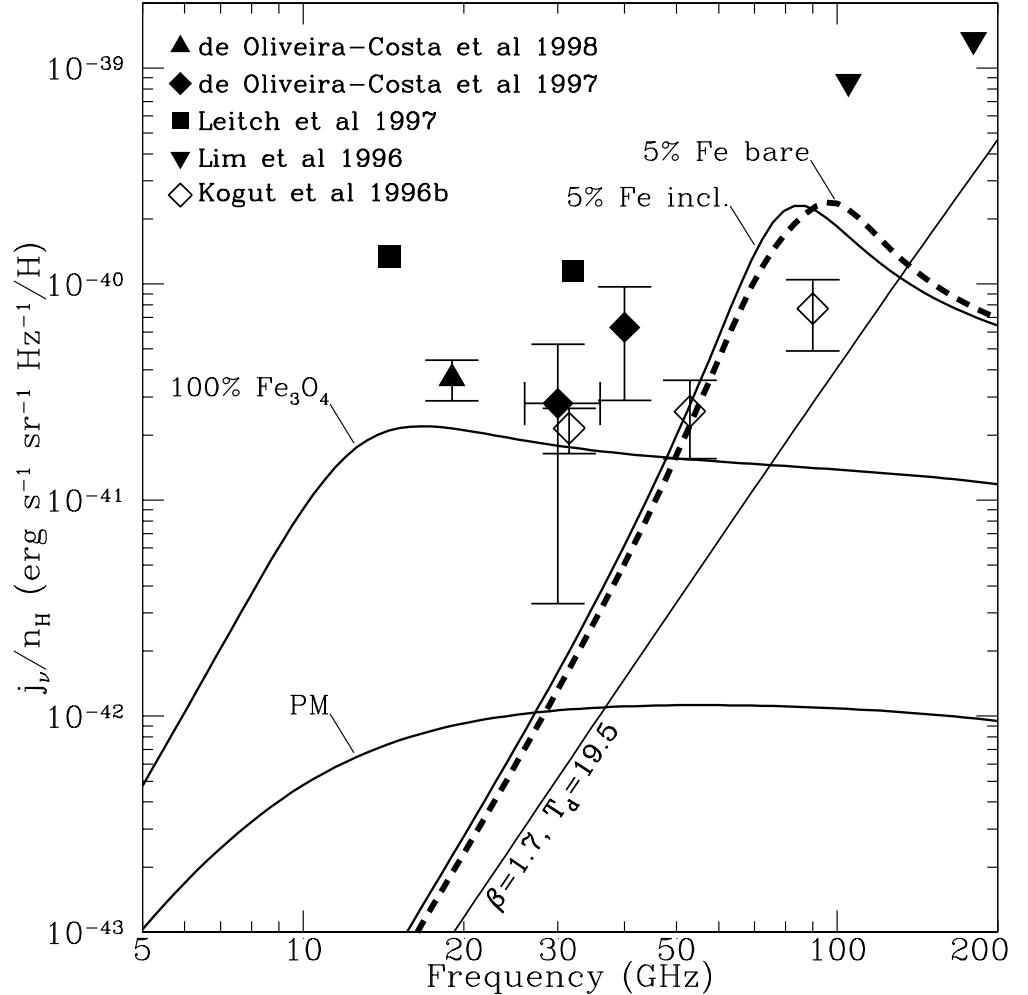}
\caption{\small%
	Thermal magnetic dipole emission from possible magnetic
	grain materials, after Draine \& Lazarian (1999).
	}
\label{fig:magdipole}
\end{figure}
Magnetic dipole emission is marginal at optical and infrared frequencies
($\nu>10^{12}$~Hz) but gets quite appreciable as the frequency of
oscillating magnetic field approaches the precession frequency of
an electron spin in the magnetic field of its neighbors, which is
$\sim 10^{10}$~Hz. The magneto-dipole microwave emissivity is enhanced
if the grain material is strongly magnetic, 
e.g. ferro- or ferrimagnetic.\footnote{%
	Another way to see that the magnetic response of
	ferromagnetic materials will dominate
	in the microwave range is through the Kramers-Kronig relations. 
	If the zero frequency magnetic and electric susceptibilities of a 
	material
	are comparable while the high frequency magnetic response is 
	negligible, the
	Kramers-Kronig relations require that the magneto-dipole adsorption
	should dominate at low frequencies (see DL99).
	}

Iron is the fifth most abundant element by mass (after H, He, O, and C),
and it is well known from absorption line studies 
that interstellar gas-phase Fe is heavily ``depleted'', with most of the
Fe locked up in dust grains (see the review by Savage \& Sembach 1996).
If $\sim$30\% of the grain mass is carbonaceous, Fe and Ni contribute
$\sim 30$\% of the mass of the remaining grain material. 

The chemical form in which the solid-phase Fe resides is not yet known,
but some fraction might be in strongly magnetic
materials, such as metallic Fe or magnetite (Fe$_3$O$_4$).
Draine \& Lazarian (1999, hereafter DL99) 
discuss emission and absorption of radiation
by grains containing magnetic materials, and show that magnetic dipole
emission due to thermal fluctuations in the magnetization can dominate
the thermal emission at microwave frequencies.

\begin{figure}[t]
\plotone{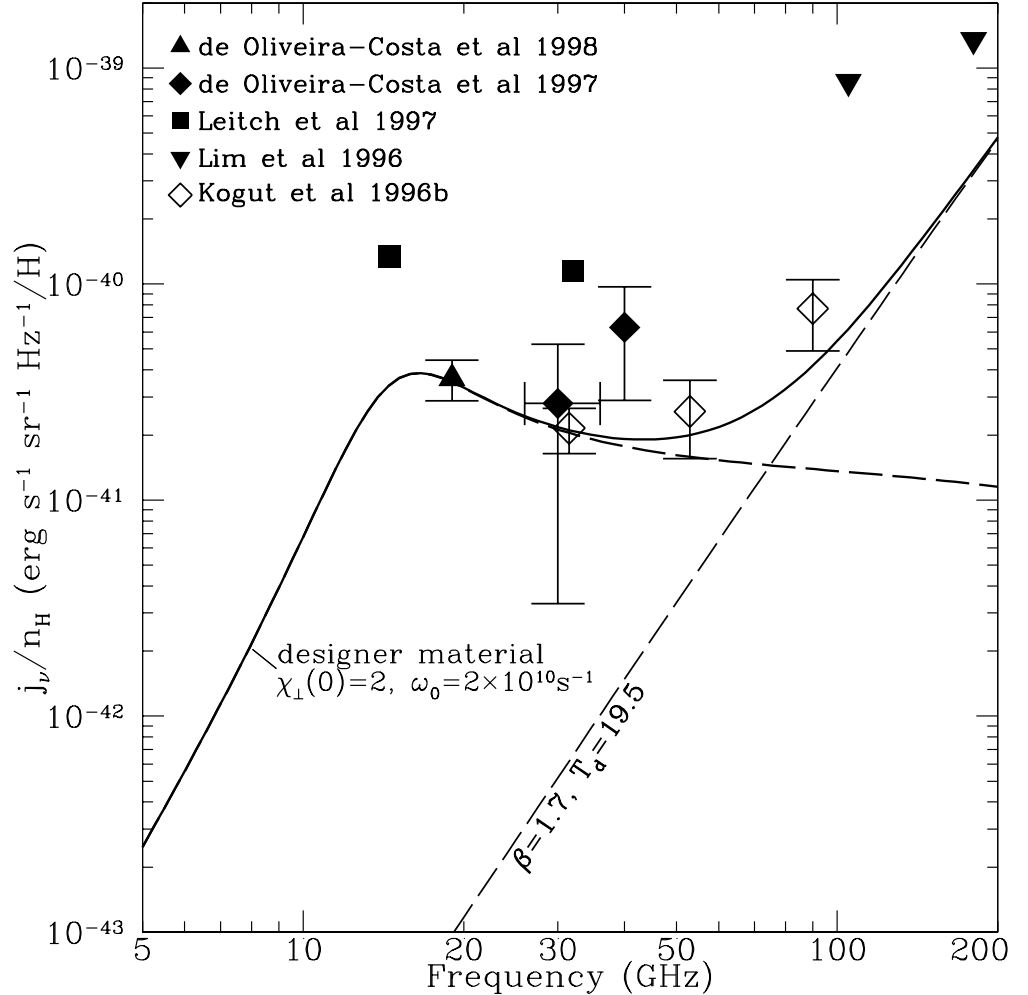}
\caption{\small%
	Thermal magnetic dipole emission from hypothetical magnetic
	grain material.
	}
\label{fig:designer}
\end{figure}

``Ordinary'' paramagnetism will result in more emission at
$\nu < 35$GHz than predicted by eq.\ (\ref{eq:betafit}), but still far
less than observed.
In order for magnetic dipole emission to reach intensities comparable
to those observed, it is necessary for a substantial fraction of interstellar
Fe to be in ferromagnetic (e.g., metallic Fe) or ferrimagnetic 
(e.g.,  magnetite Fe$_3$O$_4$) materials.
Figure \ref{fig:magdipole} shows that if 100\% of the Fe is in magnetite,
the magnetic dipole emission would approach that observed; if
as little as 5\% of the Fe is in pure metallic form -- either as bare
single-domain particles, or single-domain inclusions within larger
paramagnetic grains), the magnetic dipole emission will peak at $\sim$90GHz,
as the result of a magnetic dipole ``Fr\"ohlich resonance'' (DL99).
Existing observations show no evidence for such an emission feature, so
DL99 conclude that not more than
$\sim$5\% of the interstellar Fe can be in (pure) metallic iron.

On the other hand, the Fe could be in some
impure metallic form which is more strongly magnetic than magnetite,
but less magnetic than pure Fe metal.
If we allow ourselves the freedom to 
``tune'' the frequency-dependent magnetic susceptibility, DL99 show that
there are parameter combinations which come close to reproducing the
observed emission; an example of such a ``designer'' material is shown
in Figure \ref{fig:designer}.
The magneto-dipole contribution to the ``anomalous'' 
emission can be established via observations (see \S \ref{sec:tests}).

\section{Emission from Spinning Ultrasmall Dust Grains? Yes! 
	\label{sec:rotational}}

\begin{figure}[t]
\plotfiddle{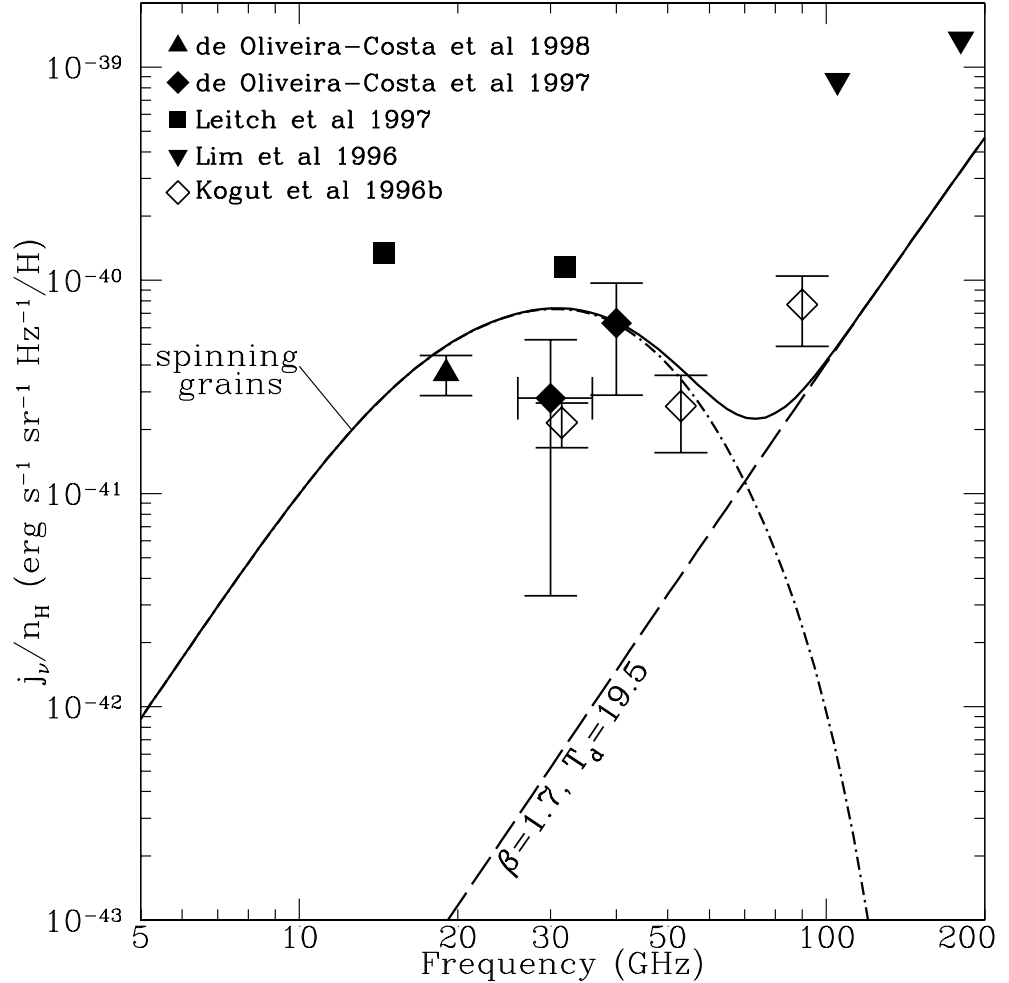}{12.4cm}{0}{65}{65}{-200}{-103}
\caption{\small%
	Predicted emission due to spinning dust grains,
	plus the extrapolated low-frequency tail of the ``electric dipole
	vibrational'' emission.
	After Draine \& Lazarian (1998b).
	}
\label{fig:spinning}
\end{figure}
The possibility of rotational emission from dust grains was apparently
first discussed by Erickson (1957).
However, as the rotational emission is proportional to the
fourth power of the angular velocity, only extremely small grains 
rotate sufficiently rapidly to produce appreciable emission. As the
evidence for such grains did not exist at the time,
the idea was not pursued. 
More recently, after the discovery
of the population of ultrasmall grains,
Ferrara \& Dettmar (1994) noted that the rotational emission from
these grains
could be observable, and DL98a
proposed that the observed ``anomalous'' emission was just the expected
rotational emission from the population of ultrasmall grains.

The rotational dynamics of dust grains, and resulting rotational emission,
are discussed by Draine \& Lazarian (1998b; herafter DL98b).
To predict expected levels of microwave emissivity DL98b had to
estimate the distribution of ultrasmall grains, their rates of rotation
and their dipole moment. An important finding of DL98b is that for
reasonable parameter values the rotational radiation from ultrasmall grains 
can account
for the observed anomalous emission.

For a population of small grains containing 5\% of the cosmic carbon abundance
(with 3\% of the cosmic carbon abundance in grains with radii 
$a < 6$\AA, or $N < 120$ C atoms), the predicted microwave emissivity of dust
in HI clouds ($n_{\rm H}=30{\rm cm}^{-3}$,
$T_{\rm gas}=100$K) is shown in Figure \ref{fig:spinning}.
We see that this rotational emission appears to be quite capable
of explaining the observed $\nu < 50$GHz ``anomalous'' emission.

A number of different processes (including collisions with ions,
collisions with neutrals, ``plasma drag'', emission of infrared photons,
and electric dipole radiation) play a role in the excitation and
damping of grain rotation.
Quite unexpectedly, DL98b found that 
collisions with ions and plasma interactions
are extremely important for the dynamics of ultrasmall
grains even in mostly neutral media. 
Because the interstellar medium is far from thermodynamic equilibrium,
the mean rotational kinetic energy can be larger or smaller than the
$1.5kT_{\rm gas}$ which would apply if the system were in LTE.
For example, $a=6$\AA\ grains
in HI clouds are estimated to
have an rms rotation rate of 20 GHz,
corresponding to a mean rotational kinetic energy of only $1.1kT_{\rm gas}$
(DL98b).
The subthermal rotation is the result of the strong damping due to
electric dipole radiation.
For ultrasmall grains quantum effects become important. DL98b found that
they limit the ``rotational drag'' on ultrasmall grains due to the
plasma.

The spectrum shown in Figure \ref{fig:spinning} is only approximate.
DL98a,b considered a number of models 
corresponding to various different assumptions
regarding electric dipole moments, grain geometry, and numbers of small
grains.
The resulting emissivities
vary by a factor of several. Figure \ref{fig:spinning} corresponds
to the most likely choice of the parameters. 

The calculations in DL98b assumed a Maxwellian form for the
distribution of rotational velocities. 
For very small grains, the actual distribution of rotational
velocities is expected to be highly non-Maxwellian, particularly
because the rotational excitation is dominated by collisions with
ions, which deliver angular impulses which can be large compared to the
rms angular momentum of the grain (DL98b).
A more detailed study of the rotational distribution function for ultrasmall
grains, and the resulting rotational emission spectrum, is in progress
(Draine \& Li 1999).

\section{Emissivity Variations \label{sec:space}}

For microwave background experiments it is important to measure the signal
as a function of the angular scale. Variations of microwave
emission from dust on scales of a few degrees will interfere with 
measurements of the cosmic background radiation on these scales.

The spatial structure of the 100$\mu$m cirrus has been studied by
Gautier {\it et al.}\ (1992), and Herbstmeier {\it et al.}\ (1998).
For measurements made with a telescope beam of angular resolution
$\theta$, the rms variations in beam-averaged surface brightness
are empirically given by
\begin{equation}
\frac{\langle(\Delta I_\nu)^2\rangle^{1/2}}{\langle I_\nu\rangle}
\approx
0.08 \left(\frac{\theta}{{\rm deg}}\right)^{0.5} 
\left(\frac{\langle I_{100}\rangle}{{\rm M\,Jy\,sr^{-1}}}\right)^{0.5} ~~.
\label{eq:fluct}
\end{equation}
where $\langle I_{100}\rangle$ is the 100$\mu$m surface brightness in that
region of the sky.
For the ``north polar cloud'' near the North Celestial Pole studied by
Gautier {\it et al.}, $\langle I_{100}\rangle=4.7 {\rm M\,Jy\,sr^{-1}}$,
and $\langle(\Delta I_\nu)^2\rangle^{1/2}/\langle I_\nu\rangle \approx
0.17 (\theta/{\rm deg})^{0.5}$.

\begin{figure}[t]
\plotfiddle{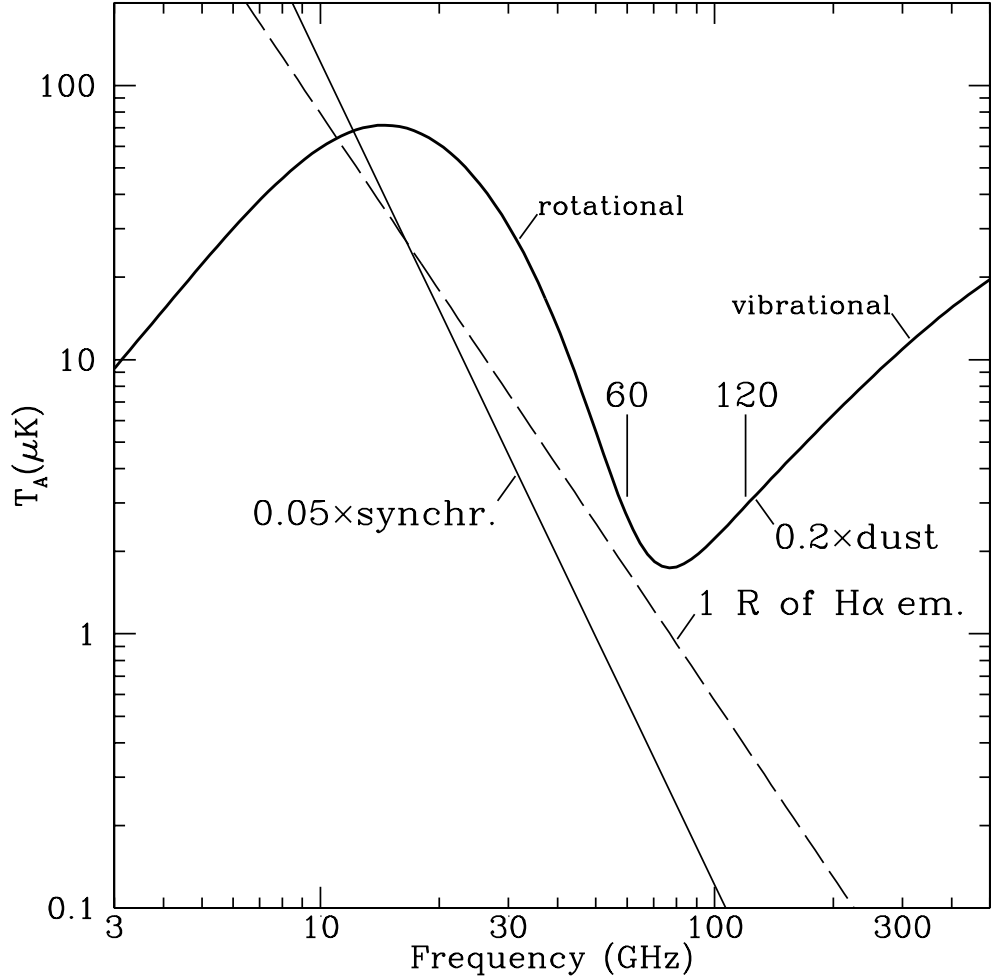}{12.4cm}{0}{65}{65}{-200}{-103}
\caption{\small%
	Estimated rms variations on angular scales of $\sim$1$^\circ$
	at intermediate galactic latitudes, $b\approx30^\circ$.
	The foreground signal is minimized in the 60--120 GHz region,
	indicated in the figure.
	After Draine \& Lazarian (1999a).
		}
\label{fig:space}
\end{figure}

Figure \ref{fig:space} 
illustrates the relative importance of the emission from dust for intermediate
Galactic latitudes, with
$I_{100}\approx 5 {\rm MJy\,sr^{-1}}$.
The rms variations of dust column density are taken
to be $20$\% on $\sim$1$^\circ$ scales, of the order expected from
eq.\ (\ref{eq:fluct}). 
The fluctuations in the H$\alpha$ emission on $\sim$1$^\circ$ scales
are taken to be 1 Rayleigh, and the accompanying free-free emission is
shown for $T_{\rm gas}\approx8000$K.
The synchrotron background is 
smoother than dust and the corresponding fluctuations are taken to be
5\% on the same angular scale.
From the standpoint of minimizing confusion
with non-CBR foregrounds, 60-120~GHz appears to be the optimal frequency
window.

Substantial variations of microwave emissivity are expected when the line of
sight crosses particular regions.
Calculations in DL98b were done for
emission from spinning dust grains in 
from dark clouds, reflection nebulae, photo-dissociation
regions etc. These variations can be used to test the spinning dust
model.

\section{Polarization \label{sec:polarization}}

Polarization of cosmic microwave background can provide valuable
insight to the physics of early Universe (Zaldarriaga 1997, Seljak
\& Zaldarriaga 1997) and therefore is the subject of intensive
theoretical studies (see Kamionkowski \& Kosowsky 1998).
To study polarization of cosmological origin one has to separate
it from foreground polarization caused by dust.

The far-infrared and sub-mm thermal ``electric dipole vibrational''
emission is expected to be linearly polarized with electric vector
${\bf E}\perp{\bf B_0}$, where ${\bf B_0}$ is the local interstellar
magnetic field.
The degree of polarization depends on the degree of alignment of
the $a > 0.1\mu$m grains which dominate both the far-infrared 
thermal emission and the polarization of starlight.
Polarizations of $\sim5\%$ are expected (Prunet {\it et al.}\ 1997)
at
frequencies ($\nu > 100$GHz) where the ``electric dipole vibrational''
emission dominates
(see more in Prunet \& Lazarian 1999).

\subsection{Spinning Grains}

At frequencies $\nu < 100$ GHz, however, we expect the spinning grains to
dominate the emission.
For perfectly aligned rotating grains the
degree of polarization would approach 100\%.
Such polarized emission could completely mask the
polarization of cosmological origin. 

Finding the actual decree of alignment for ultrasmall grains is a challenging
theoretical problem. 
The existing theories of alignment (see list of
the processes in Lazarian, Goodman \& Myers 1997) deal with
$a > 0.1\mu$m grains, while processes that can align ultrasmall grains
have not been studied until very recently.

Lazarian \& Draine (1997, 1999, henceforth LD99) concluded that paramagnetic 
dissipation is the most promissing process that can align ultrasmall 
grains over substantial ISM regions. Paramagnetic dissipation is the
most natural mechanism to consider as even ultrasmall 
grains almost certainly 
contain atoms and ions with unpaired electrons, and they will be
rotating in the galactic magnetic field.   

The effect of the galactic magnetic field ${\bf B_0}$ is to damp the
component of angular momentum perpendicular to the field, thereby bringing
about partial alignment of the angular momentum ${\bf J}$ with ${\bf B}_0$.
As a result, the electric dipole radiation from the spinning grains will
be partially linearly polarized with the 
electric vector ${\bf E}\perp{\bf B_0}$.

LD99 claim that paramagnetic relaxation for rapidly rotating grains
is more efficient than Davis-Greenstein (1951) theory predicts.
The latter implicitly assumes that the relaxation does not depend
on whether the grain rotates in stationary magnetic field or the
magnetic field rotates around a stationary grain. However, a
rotating grain gets magnetized via the Barnett effect (Landau \& Lifshitz,
1960) 
as the grain rotation effectively removes the degeneracy between ``spin-up''
and ``spin-down'' states.
This splitting of energy levels also implies that energy dissipation will
not be suppressed by very rapid rotation rates since, in effect, the
grain rotation frequency is always resonant with the splitting of energy
levels.  LD99 term this effect ``resonance relaxation''.

Work to estimate the frequency-dependent polarization expected for
the galactic foreground is in progress but estimates of polarization
of the order a few percent were obtained in LD99 for frequencies less than
30~GHz. For frequencies higher than 40~GHz the polarization becomes
negligible as the result of efficient radiative damping. We note
that the degree of grain alignment depends on the ratio of the
rotational damping time to the paramagnetic dissipation time. 
The rapid increase of rotational damping with frequency causes the decrease 
of alignment of the most rapidly rotating grains.

\subsection{Magnetic Dipole Emission}

If magnetic dipole emission 
\begin{figure}[t]
\plotfiddle{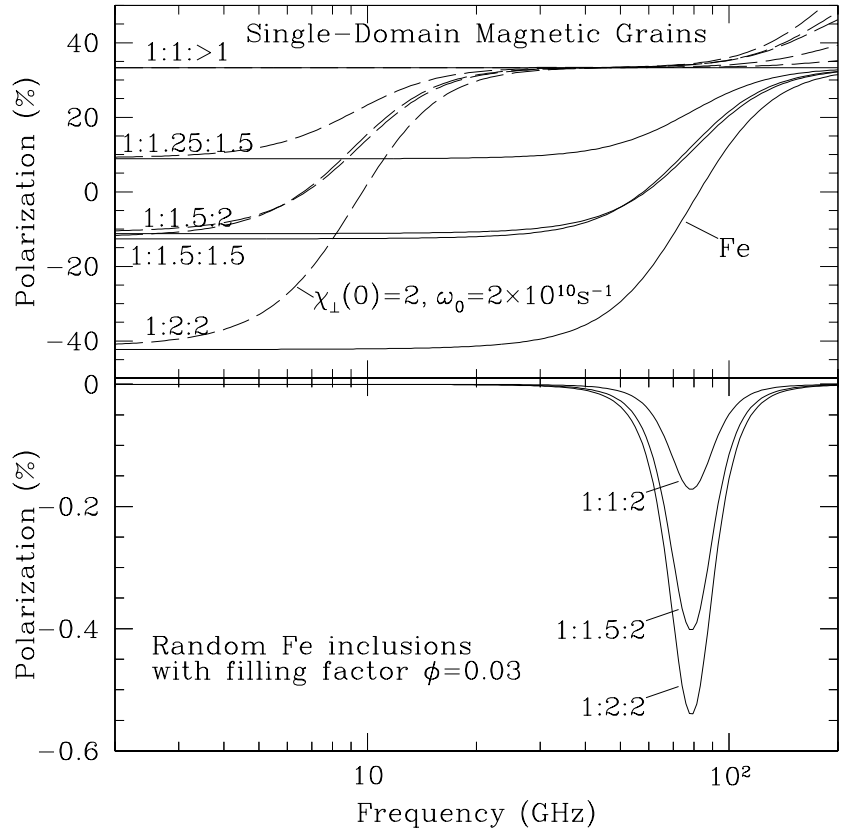}{10.6cm}{0}{55}{55}{-170}{-80}
\caption{\small%
	Upper panel: Polarization of thermal emission from 
	perfectly aligned single domain grains
	consisting of either metallic Fe (solid lines) or
	hypothetical ``designer material'' (broken lines).
	Lower panel: Polarization of microwave emission
	from perfectly aligned ellipsoids composed of material with
	metallic Fe inclusions with volume filling factor $\phi=0.03$. 
	Results are shown for various axial ratios.
	See also Figure 9 of DL99.}
\label{fig:polarization}
\end{figure}
from single-domain ferromagnetic grains
is present, it will add a distinct, frequency-dependent, polarization
signature,
with polarization of the magnetic dipole radiation changing sign close
to the frequency where the intensity of the magnetic dipole radiation
peaks (DL99).

Even if emission from rotating grains dominates the overall emission, 
the polarized emission from single-domain
ferro- and ferrimagnetic grains could be very important.
Constraining the abundance of such grains via observations is a problem
for future research.

\section{Discussion \label{sec:tests}}

Both the magnetic dipole mechanism and the spinning grain mechanism
predict that that emissivity $j_\nu$ should show a sharp decrease
as the frequency falls below $\sim15$GHz; future low-frequency
CMBR studies should be able to test this prediction.
The spectra predicted for magnetic grain materials (Figures
\ref{fig:magdipole} and \ref{fig:designer}) differ from the spectrum
predicted for spinning grains in Figure \ref{fig:spinning}, but
it must be remembered that these spectra depend on uncertain
assumptions: magnetic susceptibilities are required to estimate
magnetic dipole radiation; size distributions and electric dipole
moments are required to estimate the emission due to spinning grains.
The emission spectra in Figures \ref{fig:magdipole} -- \ref{fig:spinning}
are therefore not expected to be accurate in detail.

As pointed out by DL99, one way to distinguish between the spinning
grain mechanism and magnetic dipole emission would be to measure the
microwave emission from dust in a dense globule.
Studies of wavelength-dependent extinction (see the review by 
Mathis 1990) indicate that small
grains are underabundant in dense regions, so the rotational
emission per H should be lower in dense clouds.
The magnetic dipole emission, on the other hand, is proportional to the
total grain volume, but insensitive to the particle size.
Hence the coagulation which presumably acts to reduce the
population of ultrasmall grains in dark, dense regions will not
affect the magnetic dipole emission.
Measurement of the  microwave emission spectrum from a dark
globule can therefore reveal whether the anomalous emission is due to
spinning grains or magnetic grains.

\section{Summary \label{sec:summary}}
The principal points discussed above are as follows:
\begin{itemize}
\item There is strong evidence for 10--60 GHz emission correlated with
interstellar gas beyond that due to free-free emission.

\item The strong emission from interstellar dust at $\lambda < 50\mu$m
(see Figure \ref{fig:obs}) requires a large population of ultrasmall
dust grains.
\item These ultrasmall dust grains will rotate rapidly, and will
radiate electric dipole radiation at microwave frequencies 
(see Figure \ref{fig:spinning}) which
could explain the observed ``anomalous'' emission.
\item The preferred range of cosmic microwave background observations
is from 60 to 120~GHz.
\item The small grains will be partially aligned, with their
angular momenta tending to be either parallel or antiparallel to 
the local interstellar magnetic field ${\bf B}_0$ (Lazarian \& Draine 1997).
The electric dipole emission will therefore tend to be polarized
with ${\bf E} \perp {\bf B_0}$.
\item Polarization from rotating grains is expected to be
negligible for frequencies larger than 40~GHz.
\item If grains consist in part of magnetic material, such as metallic
Fe or magnetite Fe$_3$O$_4$, the thermal fluctuations in the magnetization
will result in strong magnetic dipole emission.  If most of the interstellar
Fe resides in a magnetic material such as magnetite, a substantial
fraction of the observed ``anomalous'' emission could be thermal
radiation from such magnetic materials.
\item If single-domain ferromagnetic grains are present and aligned, 
the magnetic dipole radiation from them will have an unusual polarization
signature.
\item Measurement of the emission from dark globules may allow us to
decide between spinning grains and magnetic dipole emission as the
dominant contribution to the anomalous emission.
\end{itemize}
\acknowledgments
We thank D.P. Finkbeiner, A. Kogut, E.M. Leitch, and
D.J. Schlegel for valuable discussions,
and R.H. Lupton for availability of the SM plotting software
used in this work.
This research was supported in part by National Science Foundation
grant AST-9619429,
and NASA grant NAG5-7030. 
AL acknowledges a Senior Research Fellowship at CITA.

\end{document}